\documentclass{mathincs}

\usepackage{amsmath}
\usepackage{epstopdf}

 \theoremstyle{definition}
 
 \theoremstyle{remark}

 \numberwithin{equation}{section}

\begin{document}
%
%
%
\title[The construction of averaged planetary motion theory by means of CAS Piranha]
{The construction of averaged planetary motion theory by means of computer algebra system Piranha}
\author[Perminov]{Perminov A.S.}

\address{%
Ural Federal University\\
51 Lenin Avenue\\
Yekaterinburg\\
Russia}

\email{perminov12@yandex.ru}

\thanks{This work is funded by RFBR according to the research project no. 18-32-00283 and the Government of the Russian Federation (Act no. 211, agreement no. 02.A03.21.0006).}
\author[Kuznetsov]{Kuznetsov E.D.}

\address{%
Ural Federal University\\
51 Lenin Avenue\\
Yekaterinburg\\
Russia}

\email{Eduard.Kuznetsov@urfu.ru}


\date{November 30, 2017}
\dedicatory{}

\begin{abstract}

The application of computer algebra system Piranha to the investigation of the planetary problem is described in this work. Piranha is an echeloned Poisson series processor authored by F.~Biscani from Max Planck Institute for Astronomy in Heidelberg. Using Piranha the averaged semi-analytical motion theory of four-planetary system is constructed up to the second degree of planetary masses. In this work we use the algorithm of the Hamiltonian expansion into the Poisson series in only orbital elements without other variables. The motion equations are obtained analytically in time-averaged elements by Hori-Deprit method. Piranha showed high-performance of analytical manipulations. Different properties of obtained series are discussed. The numerical integration of the motion equations is performed by Everhart method for the Solar system's giant-planets and some exoplanetary systems.

\end{abstract}

\maketitle

\section{Introduction}

The investigation of planetary systems orbital evolution, in particular the Solar system, is one of important problems of celestial mechanics. This study is closely related to the problem of the stability of the planetary system. It is important because according to modern ideas only in stable systems the life is possible to the origin.

In XVIII and XIX centuries J.L.~Lagrange~\cite{Lagrange1781, Lagrange1782} and P.S.~Laplace~\cite{Laplace1829} made a significant contribution to the problem of the stability of the Solar system. It was found that the semi-major axes of the planetary orbits have not the secular perturbations in non-resonant case of the motion theory of the first order in the masses (it is well-known Laplace--Lagrange theorem). So, it is proved analytically that the planetary motion is quasi-periodically in this case.

In XX century A.N.~Kolmogorov, V.I.~Arnold and Yu.~Moser created KAM-theory for the study of the influence of small perturbations on the dynamics of weakly perturbed Hamiltonian systems. The question of the Solar system stability becomes an important area of applicability of this theory. From the point of view of convergence for the series of the motion equations, even for very small perturbations, the points of convergence do not fill the phase space everywhere dense. Consequently, the measure of the points of divergence becomes non zero. For such trajectories in the phase space, it is impossible to predict the behaviour of the system at exponentially large times.

In the 1960's G.~Hori and A.~Depri independently proposed the new averaged method based on Lee transformations theory and called the Hori-Depri method. In the comparison with the other averaging methods it is more simple and explicitly for the implementation.

The essence of any averaging method is the excluding from the motion equations all short-periodic terms given by fast variables (mean anomalies or mean longitudes in the planetary motion). The transformations between the osculating variables and their averaged values is given by the function for the change of variables. The periodicity of the functions for the change with respect to the fast variables provides the closeness of the old (osculating) and the new (averaged) sets of variables. We chose this method for the construction of our averaged planetary motion theory.

It is a brief overview of the works on the study of the Solar system orbital evolution. The secular orbital evolution of 8 planets of the Solar system is numerically studed in work of J.~Laskar~\cite{Laskar1988} over 30~million years and it is studed in works of J.~Applegate \cite{Applegate1986} and J.~Laskar~\cite{Laskar1989, Laskar1990} over 200~million years. The chaotic properties of the Solar system, the estimations of the size of the chaotic zones and the chaotic diffusion of the planets are investigated in the set works of J.~Laskar~\cite{Laskar1994, Laskar1996, Laskar2008}. The dynamical stability of the Solar system over times more than few billion years is studed, for example, by K.~Batygin~\cite{Batygin2008}, T.~Ito and K.~Tanikawa~\cite{Ito2002} and J.~Laskar~\cite{Laskar2008}.

The application of KAM-theory is considered by A.~Giorgilli, M.~Sansottera and U.~Locatelli to the investigation of the stability of the Sun~-- Jupiter~-- Saturn system in~\cite{Giorgilli2009} and to the planar model of the Sun~-- Jupiter~-- Saturn~-- Uranus system in~\cite{Giorgilli2011, Giorgilli2017, Sansottera2013}.

The construction of the semi-analytical motion theory and the investigation of the dynamical evolution and the stability of the Sun~-- Jupiter~-- Saturn system over 10~billion years is considered by E.D.~Kuznetsov and K.V.~Kholshevnikov in the next set of works~\cite{Kholshevnikov2001, Kholshevnikov2002, Kuznetsov2004, Kuznetsov2006, Kuznetsov2009}.

In this moment by the investigations of above mentioned authors the following ideas about the dynamic properties of the Solar system on long time intervals are formed. It is fully reviewed in~\cite{Kholshevnikov2007}. The planetary motion in the Solar system is quasi-periodic on time scales of about $10^6-10^7$ years. In this case the behaviour of orbital elements for all planets of the Solar system is fully predictable. The planetary motion remains quasi-periodic on time scales of about $10^8-10^9$ years, but the orbital elements are predictable for only giant planets. The motion of the inner planets begins to become the chaotic. For example, Mercury can be ejected from the Solar system in several billion years as it is shown in~~\cite{Laskar1994}. The motion of the giant planets is almost quasi-periodic, but the motion of the inner planets is completely chaotic on time scales of about $10^{10}$ years. In this case the information about the phase of the planetary motion is unknown.

The existence of the initial conditions that lead to both regular and chaotic solutions for the planetary motion in the Solar system is shown in~\cite{Guzzo2005, Guzzo2006, Murray1999a}.

\section{The motion theory}

Our main scientific objective is the construction of semi-analytical motion theory of four-planetary problem. In order to investigate planetary systems dynamical evolution on long-time intervals we need to obtain motion equations in time-averaged orbital elements. The using of these elements allows to eliminate short-periodic perturbations in the planetary motion. Further these equations can be numerically integrated for the studying of the orbital evolution of various planetary systems, such as the Solar system's giant planets and some exoplanetary systems.

The first stage of our work is the expansion of the planetary system Hamiltonian into the Poisson series in all orbital elements. The algorithm of this expansion   is described in detail in~\cite{Perminov2015}. Let us briefly consider it here. The Hamiltonian of the four-planetary problem is written in Jacobi coordinates~\cite{Murray1999b}. It is the hierarchical coordinate system in which the position of each following body is determined relative to the barycentre of the previously including bodies set. The Hamiltonian in Jacobi coordinates is written in more simple form and shown here

\begin{equation} \label{eq1}
	h=-\sum^{4}_{i=1}\frac{M_i \kappa_i^2}{2a_i}+\mu\;Gm_0\Bigl(\sum^4_{i=2}\frac{m_i (2\mathbf{r}_i\mathbf{R}_i+\mu R_i^2)}{r_i \tilde{R}_i (r_i+\tilde{R}_i)}-\sum^4_{i=1}\sum^{i-1}_{j=1}\frac{m_i m_j}{|\mathbf{\rho}_i - \mathbf{\rho}_j|}\Bigr),
\end{equation}
where
\begin{equation*}
	\mathbf{R}_i=\sum^i_{k=1}\frac{m_k}{\bar{m}_k}\mathbf{r}_k,\quad\tilde{R}_i=\sqrt{r^2_i+2\mu\mathbf{r}_i\mathbf{R}_i+\mu^2 R^2_i},\quad|\mathbf{\rho}_i - \mathbf{\rho}_j|=\mathbf{r}_i-\mathbf{r}_j+\mu\sum^{i-1}_{k=j}\frac{m_k}{\bar{m}_k}\mathbf{r}_k.
\end{equation*}
Here $1\leq{j}<i\leq{4}$, $\rho_k$ and $\mathbf{r}_k$~are the barycentric and Jacobi radius vectors of \textit{k}-th planet, $\mu m_k$ is the mass of the planet in the star mass $m_0$, $\bar{m}_k=1+\mu m_1+\ldots +\mu m_k$, $M_i=m_i\bar{m}_{i-1}/\bar{m}_i$, $\kappa_i^2=Gm_0\bar{m}_i/\bar{m}_{i-1}$. The small parameter $\mu$ is equal to the ratio of the sum of planetary masses to the mass of the star and will play the role of the expansion parameter.

The first sum in~(\ref{eq1}) is the undisturbed Hamiltonian. The expression in the figure brackets of~(\ref{eq1}) is the disturbing function, which describes gravitational interaction between planets. The double sum is the main part of the disturbing function and the single sum is so-called the second part. We need to expand the disturbing function into the Poisson series in all orbital elements and the small parameter. The second system of the Poincare elements is used for the construction of the Hamiltonian expansion. This system has only one angular element~-- mean longitude, that allows to simplify an angular part of the expansion. The Poincare elements are defined through the Keplerian elements as shown here~\cite{Sharlier1927}
\begin{eqnarray}
	L_i=M_i\sqrt{\kappa_i^2a_i},\;\;\; \lambda_i=\Omega_i+\omega_i+l_i, \nonumber\\
	\xi_{1i}=\sqrt{2L_i(1-\sqrt{1-e_i^2})}\cos{(\Omega_i+\omega_i)},\;
	\xi_{2i}=\sqrt{2L_i\sqrt{1-e_i^2}(1-\cos{I_i})}\cos{\Omega_i}, \nonumber\\
	\eta_{1i}=-\sqrt{2L_i(1-\sqrt{1-e_i^2})}\sin{(\Omega_i+\omega_i)},\;
	\eta_{2i}=-\sqrt{2L_i\sqrt{1-e_i^2}(1-\cos{I_i})}\sin{\Omega_i}, \nonumber
\end{eqnarray}
where $a_i$~-- semi-major axis of an orbital ellipse, $e_i$~-- eccentricity of the orbit, $I_i$~-- inclination of the orbit, $\Omega_i$~-- longitude of the ascending node, $\omega_i$~-- argument of the pericenter, $l_i$~-- mean anomaly of planet $i$. The Poincare elements are canonical elements and three pairs of these are canonical conjugated as the momentum and its corresponding coordinate, namely $L$ and $\lambda$, $\xi_1$ and $\eta_1$, $\xi_2$ and $\eta_2$. Note that $\xi_{1i}$, $\eta_{1i} \propto e_i$ (so-called the eccentric elements) and $\xi_{2i}$, $\eta_{2i} \propto I_i$ (the oblique elements).

The expansion of the main part of the disturbing function is shown here up to the third degree of the small parameter
\begin{equation} \label{eq2}
	\frac{1}{|\mathbf{\rho}_i - \mathbf{\rho}_j|}=\frac{1}{\Delta_{ij}}-\mu\frac{A_{ij}}{\Delta_{ij}^3}+\mu^2\Bigl(\frac{3}{2}\frac{A_{ij}^2}{\Delta_{ij}^5}-\frac{1}{2}\frac{B_{ij}}{\Delta_{ij}^3}\Bigr)+\mu^3\Bigl(-\frac{5}{2}\frac{A_{ij}^3}{\Delta_{ij}^7}+\frac{3}{2}\frac{A_{ij}B_{ij}}{\Delta_{ij}^5}\Bigr)+\dots.
\end{equation}

The inverse distance of the Jacobi radius vectors $1/\Delta_{ij}=|\mathbf{r}_i - \mathbf{r}_j|^{-1}$ in (\ref{eq2}) is expanded into the series in the Legendre polynomials
\begin{equation} \label{eq3}
	\frac{1}{\Delta_{ij}}=\frac{1}{r_j}\Bigl(1+\Bigl(\frac{r_i}{r_j}\Bigr)^2-2\Bigl(\frac{r_i}{r_j}\Bigr)\cos{H}\Bigr)^{-\frac{1}{2}}=\frac{1}{r_j}\sum^\infty_{n=0}\Bigl(\frac{r_i}{r_j}\Bigr)^n P_n(\cos{\theta_{ij}}),
\end{equation}
where $P_n$~--~the Legendre polynomial of $n$-th degree, $\theta_{ij}$~--~an angle between radius vectors $\mathbf{r}_i$ and $\mathbf{r}_j$. The series expansion for $1/\Delta_{ij}$ in the Legendre Polynomials is the intermediate step of the expansion algorithm. Further we express the Legendre polynomials through angles $\theta_{ij}$. Then angles $\theta_{ij}$ are expanded into the series through eccentric and oblique Poincare elements. So we have expressed the quantities $1/\Delta_{ij}$ in the Poincare elements. The analysis of the convergence for the quantities $1/\Delta_{ij}$ depending on the maximum included degree of Legendre polynomials are given in \cite{Perminov2015}.

The second part of the disturbing function is expanded into the series as shown here
\begin{align} \label{eq4}
	\frac{2\mathbf{r}_i\mathbf{R}_i+\mu\mathbf{R}_i^2}{r_i \tilde{R}_i(r_i+\tilde{R}_i)}=\frac{C_i}{r_i^3}+\mu\Bigl(-\frac{3}{2}\frac{C_i^2}{r_i^5}+\frac{1}{2}\frac{D_i}{r_i^3}\Bigr)+&\mu^2\Bigl(\frac{5}{2}\frac{C_i^3}{r_i^7}-\frac{3}{2}\frac{C_i D_i}{r_i^5}\Bigr)\\\nonumber
	+&\mu^3\Bigl(-\frac{35}{8}\frac{C_i^4}{r_i^9}+\frac{15}{4}\frac{C_i^2D_i}{r_i^7}-\frac{3}{8}\frac{D_i^2}{r_i^5}\Bigr)+\dots.
\end{align}

The quantities $A_{ij}$, $B_{ij}$, $C_i$, $D_i$ are introduced for simplicity and defined as following way
\begin{equation} \label{eq5}
	A_{ij}=(\mathbf{r}_i-\mathbf{r}_j)\sum^{i-1}_{k=j}\frac{m_k}{\bar{m}_k}\mathbf{r}_k,\;B_{ij}=\Bigl(\sum^{i-1}_{k=j}\frac{m_k}{\bar{m}_k}\mathbf{r}_k\Bigr)^2,\;C_i=\mathbf{r}_i\sum^{i-1}_{k=1}\frac{m_k}{\bar{m}_k}\mathbf{r}_k,\;D_i=B_{i1}.
\end{equation}

{\bf It should be noted the following.} In previous works \cite{Perminov2015, Perminov2017} the cosines of the angles $\theta_{ij}$ between Jacobi radius vectors $\mathbf{r}_i$ and $\mathbf{r}_j$ are saved in the Hamiltonian expansion and the motion equations as the symbol variables without inner structure. Therefore the obtained series contain six such variables (for each of six possible pairs of radius vectors) in addition to the orbital elements and the mass parameters. However, it leads to necessity of the introduction of the motion equations for these cosines and complication of the integration process. For these reasons in this work the cosines represent as the series in the Poincare elements. It allows to reduce the number of used variables but the number of the terms in the constructed series (\ref{eq3}, \ref{eq5}) is increased.

The second stage of our work is the construction of the averaged Hamiltonian of the four-planetary problem  by the Hori-Deprit method. It is characterized by efficiency and very ease for the computer implementation. In more detail it is described in~\cite{Kholshevnikov1985, Perminov2016}. The variables of the problem can be divide into the slow variables $x$ and the fast $\lambda$. The corresponding averaged ones denoted as $X$ and  $\Lambda$. The rates of the change for the slow variables are proportionally to the small parameter $\mu$ while the rates of the change for the fast variables are proportionally to the mean motions of the planets. The averaging procedure is applied to the Hamiltonian in the osculating elements (\ref{eq1}) with respect to the mean longitudes~$\lambda$ to exclude short-periodic perturbations.

In that way the averaged Hamiltonian is wrote as the series in the small parameter

\begin{equation} \label{eq6}
	H(X)=H_0+\sum_{m=1}^\infty{\mu^mH_m(X)},
\end{equation}
where $H_m$ are obtained up to the second order by using of the main equation of the Hori--Deprit method
\begin{align*}
	&H_0=h_0,\quad H_1=\{T_1,h_0\}+h_1,\\
	&H_2=\{T_2,h_0\}+\{T_1,h_1\}+\frac{1}{2}\{T_1,\{T_1,h_0\}\}.
\end{align*}
The figure brackets denotes the Poisson brackets with respect to the Poincare elements. Term $h_0$~is the undisturbed Hamiltonian, $h_1$ is the disturbing function of the Hamiltonian in the osculating elements, $T_1$ and $T_2$ are terms of the generating function for the transformation between the osculating and the averaged orbital elements. The generating function is defined from the main equation of the Hori--Deprit method also. See~\cite{Kholshevnikov1985} in more detail.

The averaged motion equations for $X$ and $\Lambda$ are obtained by using of Poisson brackets
\begin{equation} \label{eq7}
	\frac{\mathrm{d}X}{\mathrm{d}t}=\{H,X\},\;\;\;\frac{\mathrm{d}\Lambda}{\mathrm{d}t}=\{H,\Lambda\}.
\end{equation}
The averaged elements $X, \Lambda$ are given by using of the functions for the change of variables $u_m$, $v_m$
\begin{equation} \label{eq8}
	X=x+\sum_{m=1}^\infty{(-1)^m\mu^mu_m(x,\lambda)},\;\;\;\Lambda=\lambda+\sum_{m=1}^\infty{(-1)^m\mu^mv_m(x,\lambda)},
\end{equation}
where these functions of the first and the second orders are defined by the following way
\begin{equation*}
	u_1=\{T_1,X\},\;u_2=\{T_2,X\}+\frac{1}{2}\{T_1,\{T_1,X\}\},\quad v_1=\{T_1,\Lambda\},\;v_2=\{T_2,\Lambda\}+\frac{1}{2}\{T_1,\{T_1,\Lambda\}\}.
\end{equation*}

The Python's program code which is used for the construction of the Hamiltonian expansion and the motion equations according to previously mentioned algorithms is available on GitHub repository https://github.com/celesmec/eps.

\section{Overview of Piranha}

All analytical manipulations are implemented by means of computer algebra system Piranha~\cite{Biscani2017}. Piranha is an echeloned Poisson series processor for analytical manipulations with different series types. The author of this program is Francesco Biscani from Max Planck Institute for Astronomy in Heidelberg, Germany. The algorithms for the analytical transformations can be implemented as Python's executable scripts. Piranha is a special system for the application in celestial mechanics and it is showed a high-performance in the process of computations. The comparison of the performance between one of Piranha's old version and other algebraic manipulators is presented in~\cite{Biscani2009}.

Calculations were performed on Six-core PC with 3600~MHz Core~i7 processor and 128~Gb available memory. Unix-like OS Ubuntu~16 and Python~2.7 is used. Such amount of memory is required for parallel calculations of several series.

Piranha has three supported numerical types~-- integer, rational (fractions with integer numerators and denominators) and real (with different accuracy). Piranha works with the following series types~-- polynomials, Poisson series and echeloned Poisson series. All calculated series represent as echeloned Poisson series with rational coefficients. It allows to eliminate the rounding errors and provides the arbitrary precision of the resulting series.

{\bf Note that the following.} In this work we use the version of Piranha where the new algorithm of very large series multiplication is implemented. It allows sufficiently decrease the multiplication time by using of a few processor's threads. See it more detail in~\cite{Biscani2015}.

Main functionality features of Piranha are named here.

\begin{itemize}
	\item \verb|+|,~\verb|-| and \verb|*| perform the summation and the multiplication of the series.
	\item \verb|invert(arg)| provides the multiplicative inverse of series \verb|arg|.
	\item \verb|cos(x)| and \verb|sin(x)| are used for the construction of the angular part of the series.
	\item \verb|truncate_degree(arg, max_degree, names)| truncates all terms whose degrees of variables greater than \verb|max_degree| in series \verb|arg|. The argument \verb|names| is a list of names of variables which are chosen for the truncation.
	\item \verb|subs(arg, name, x)| is used for the substitution series \verb|x| in variable \verb|name| of series \verb|arg|.
	\item \verb|evaluate(arg, eval_dict)| is used for the estimation of series \verb|arg| by values from dictionary \verb|eval_dict| which consists pairs (the variable name~-- its numerical value).
	\item \verb|save(arg, path_to_file)| saves the resulting series in a file. 
	\item \verb|load(path_to_file)| loads the series from a file.
	\item \verb|degree(arg, names)| and \verb|ldegree(arg, names)| returns the total degree/the low degree of series \verb|arg| relative to variables which named in \verb|names|.
	\item \verb|t_degree(arg, names)| and \verb|t_ldegree(arg, names)| returns the total degree/the low degree of the trigonometric part of series \verb|arg| relative to the variables which named in \verb|names|.
	\item \verb|partial(arg, name)| is the partial derivative of series \verb|arg| relative to variable \verb|name|.
	\item \verb|integrate(arg, name)| is the undefined integral of series \verb|arg| relative to variable \verb|name|.
	\item \verb|t_integrate(name)| calculates the time-integral of the trigonometric part of series \verb|arg|.
	\item \verb|pbracket(f, g, p_list, q_list)| computes the Poisson bracket of series \verb|f| and \verb|g| relative to the momentum named in \verb|p_list| and the coordinates named in \verb|q_list|.
	\item \verb|transformation_is_canonical(new_p, new_q, p_list, q_list)}| returns \verb|True| if the transformation between old momentum \verb|p_list|, old coordinates \verb|q_list| and new momentum \verb|new_p|, new coordinates \verb|new_q| is canonical.
\end{itemize}

\section{Poincare processor}

The new Poincare processor was implemented for the construction of the basic expansions by using of the next scripts.
\begin{itemize}
	\item \verb|xyz(n)| returns the series expansion for the Cartesian coordinates $x$, $y$, $z$.
	\item \verb|rrr(n, deg)| returns the series expansion for the distance $r$ from the barycentre of the system to the planetary position.
	\item \verb|one_r(n, deg)| returns the series expansion for the inverse distance $1/r$ from the barycentre of the system to the planetary position.
\end{itemize}
We expand the quantities $x$, $y$, $z$, $r$ and $1/r$ into the series in the Bessel functions and then the Bessel functions expressed through the Poincare elements. All above-mentioned series are constructed up to \verb|n|-th order of eccentric and oblique elements. If integer argument \verb|deg| provides more than $1$ it rises the resulting series to the power \verb|deg|. Thus we can implement the following scripts.
\begin{itemize}
	\item \verb|cosine(n, deg)| returns the series expansion for the cosine of an angle $\theta_{ij}$ between Jacobi radius vectors from barycentre to positions of planets $i$ and $j$.
	\item \verb|one_delta(n, c_max, deg)| returns the series expansion for the quantity $1/\Delta_{ij}$. The Legendre polynomials in expansion (\ref{eq3}) are expressed through cosines of angles between Jacobi radius vectors. Here \verb|c_max| is the maximum degree of cosines which are used for the construction of the expansion. Required ratios of the distances $r_i/r_j$ in (\ref{eq3}) are combined from series $r_i$ and $1/r_j$ for different planets $i$ and $j$.
\end{itemize}
Using combinations of these scripts allows us to construct the scalar products of radius vectors and quantities (\ref{eq5}). Then terms of the main part (\ref{eq2}) and the second part (\ref{eq4}) of the disturbing function can be obtained up to the given order of eccentric and oblique elements.

\section{Implementation of Hori-Deprit method}

The following our scripts are required for the full implementation of Hori-Deprit method.
\begin{itemize}
	\item \verb|ham_1()| computes the first item of the averaged Hamiltonian $H_1$ which is equal to the difference between the whole disturbing function $h_1$ and its trigonometric part $t_1$. In that way $h_1=H_1+t_1$. We can take quantity $H_1$ excluding of angular terms from $h_1$.
	\item \verb|eqn_1()| gives the first-order motion equations.
	\item \verb|int_1()| computes the first item of the generating function $T_1$ which equals to time-integrated trigonometric part of $h_1$. Therefore $T_1$ obtained by apply function \verb|t_integrate| to series $t_1$.
	\item \verb|chn_1()| gives the first-order functions for the change of variables.
	\item The Poisson bracket $\{T_1,h_1\}$ is constructed as the sum of products of the partial derivatives relative to the Poincare elements of all planets
\begin{equation*}
	\{T_1,h_1\}=\sum_{k=1}^4\Bigl(\frac{\partial T_1}{\partial L_k}\frac{\partial h_1}{\partial \lambda_k}-\frac{\partial T_1}{\partial \lambda_k}\frac{\partial h_1}{\partial L_k}+\frac{\partial T_1}{\partial \xi_{1k}}\frac{\partial h_1}{\partial \eta_{1k}}-\frac{\partial T_1}{\partial \eta_{1k}}\frac{\partial h_1}{\partial \xi_{1k}}+\frac{\partial T_1}{\partial \xi_{2k}}\frac{\partial h_1}{\partial \eta_{2k}}-\frac{\partial T_1}{\partial \eta_{2k}}\frac{\partial h_1}{\partial \xi_{2k}}\Bigr),
\end{equation*}
where partial derivatives of $T_1$ are actually the constructed functions for the change of variables of the first order. We implemented script \verb|pb_t1h1(n)| for it, where hereinafter \verb|n| is the maximum degree of eccentric and oblique elements.
	\item The Poisson bracket $\{T_1,h_0\}$ is defined by script \verb|pb_t1h0(n)| as
\begin{equation*}
\{T_1,h_0\}=\sum_{k=1}^4\omega_k\frac{\partial T_1}{\partial \lambda_k},
\end{equation*}
where $\omega_k=M_k^3\kappa_k^4/L_k^3$ is the motion frequency.
	\item The Poisson bracket $\frac{1}{2}\{T_1,\{T_1,h_0\}\}$ is constructed by script \verb|pb_t1t1h0(n)| as shown here
\begin{eqnarray*}
	\frac{1}{2}\{T_1,\{T_1,h_0\}\}=\frac{1}{2}\sum_{k=1}^4\sum_{m=1}^4\Bigl(\frac{\partial T_1}{\partial L_k}\frac{\partial^2 T_1}{\partial \lambda_k\lambda_m}\omega_m-\frac{\partial T_1}{\partial \lambda_k}\frac{\partial}{\partial L_k}\Bigl(\omega_m\frac{\partial T_1}{\partial \lambda_m}\Bigr)+\\\frac{\partial T_1}{\partial \xi_{1k}}\frac{\partial^2 T_1}{\partial \eta_{1k}\lambda_m}\omega_m-\frac{\partial T_1}{\partial \eta_{1k}}\frac{\partial^2 T_1}{\partial \xi_{1k}\lambda_m}\omega_m+\frac{\partial T_1}{\partial \xi_{2k}}\frac{\partial^2 T_1}{\partial \eta_{2k}\lambda_m}\omega_m-\frac{\partial T_1}{\partial \eta_{2k}}\frac{\partial^2 T_1}{\partial \xi_{2k}\lambda_m}\omega_m\Bigr).
\end{eqnarray*}
	\item \verb|ham_2()| computes the second item of the averaged Hamiltonian. Lets denote $\Phi_2=\{T_1,h_1\}+\frac{1}{2}\{T_1,\{T_1,h_0\}\}$. Then $H_2=\Phi_2-t_2$, where quantity $t_2$ is trigonometric part of $\Phi_2$.
	\item \verb|eqn_2()| gives the second-order motion equations.
	\item \verb|int_2()| computes the second item of the generating function $T_2$ obtained by apply function \verb|t_integrate| to series $t_2$.
	\item \verb|chn_2()| gives the second-order functions for the change of variables.
\end{itemize}

\section{Properties of the constructed series}

Lets consider some properties of the basic series and the Hamiltonian expansion. Table~\ref{tab1} presents the calculation time of the series, the number of its terms and the series truncation error for the basic series. The series truncation error is determined as the relative difference between the evaluated series expansion and the evaluated exact expression. It is calculated for the orbital elements of giant-planets of the Solar System. The series are obtained up to $5$-th degree of eccentric and oblique Poincare elements correspondingly.

\begin{table}[h]
	\caption{The properties of the basic series} \label{tab1}
	\begin{tabular}{l|cccccc}
	\hline
	series & $x$, $y$ & $z$ & $r$ & $1/r$ & $r_i/r_j$ & $\cos{\theta_{ij}}$ \\
	\hline
	number & $96$ & $116$ & $46$ & $41$ & $3\;486$ & $2\;438$ \\
	accuracy & $10^{-9}$ & $10^{-7}$ & $10^{-9}$ & $10^{-8}$ & $10^{-8}$ & $10^{-7}$ \\
	time & $0.05^\text{s}$ & $0.05^\text{s}$ & $0.15^\text{s}$ & $0.15^\text{s}$ & $0.05^\text{s}$ & $0.30^\text{s}$ \\
	\hline
	\end{tabular}
\end{table}

Table \ref{tab2} presents the number of terms of the Hamiltonian items and its calculation time. In table head we using of the following designations~-- $n_k$~are maximum degrees of eccentric and oblique Poincare elements in terms with $\mu^k$, $c_k$~are maximum used powers of the cosines of angles in these terms, $N_k$~is the number of the terms with $\mu^k$, $t$~is the calculation time of the series expansion. The sum of the number of the terms in each groups is showed in the bottom of the table. The total number of terms in the Hamiltonian is $6\;458\;615$ (included $4$ terms of the undisturbed Hamiltonian).

\begin{table}[h]
	\caption{The properties of the Hamiltonian expansion} \label{tab2}
	\begin{tabular}{c|ccc|ccc|ccc}
	\hline
	Indices & $n_1,c_1$ & $N_1$ & $t$ & $n_2,c_2$ & $N_2$ &$t$ & $n_3,c_3$ & $N_3$ &$t$ \\
	\hline\hline
	$i,j$ & \multicolumn{9}{c}{The main part of the disturbing function} \\
\hline
	$1,2$ & $5$, $25$ & $591\;376$ & $96^\text{s}$ & $3$, $15$ & $175\;786$ & $42^\text{s}$ & $3$, $15$ & $487\;648$ & $650^\text{s}$ \\
	$2,3$ & $5$, $25$ & $591\;376$ & $96^\text{s}$ & $3$, $15$ & $175\;786$ & $42^\text{s}$ & $3$, $15$ & $487\;648$ & $650^\text{s}$ \\
	$3,4$ & $5$, $25$ & $591\;376$ & $96^\text{s}$ & $3$, $15$ & $175\;786$ & $42^\text{s}$ & $3$, $15$ & $487\;648$ & $650^\text{s}$ \\
	$1,3$ & $5$, $20$ & $385\;460$ & $84^\text{s}$ & $2$, $10$ & \;\;$82\;874$ & $13^\text{s}$ & $2$, $10$ & $505\;541$ & $112^\text{s}$ \\
	$2,4$ & $5$, $20$ & $385\;460$ & $84^\text{s}$ & $2$, $10$ & \;\;$82\;874$ & $13^\text{s}$ & $2$, $10$ & $505\;541$ & $112^\text{s}$ \\
	$1,4$ & $5$, $15$ & $223\;476$ & $71^\text{s}$ & $2$, \;\;$5$ & \;\;$41\;988$ & $10^\text{s}$ & $2$, \;\;$5$ & $432\;744$ & $100^\text{s}$ \\
	\hline\hline
	$k$ & \multicolumn{9}{c}{The second part of the disturbing function} \\
	\hline
	$2,3,4$ & $5$, \;\;-- & \;\;$12\;852$ & \;\;$6^\text{s}$ & $3$, \;\;-- & \;\;\;\;$7\;994$ & $3^\text{s}$ & $3$, \;\;-- & \;\;$27\;369$ & \;\;$12^\text{s}$ \\
	\hline\hline
	\end{tabular}
	\begin{tabular}{cccccc}
	The Hamiltonian expansion & $N_1=2\;768\;524$ & & $N_2=735\;094$ & & $N_3=2\;934\;139$ \\
	\hline
	\end{tabular}
\end{table}

The approximation accuracy of the Hamiltonian by the series expansion is evaluated for giant-planets of the Solar system and two exoplanetary systems HD141399 and HD160691. All of these planetary systems have four planets. The Kepler orbital elements and masses of the Solar system's giant-planets are given from ephemeris DE430~\cite{Folkner2014} on date 01/01/2000 and correspond to the mean ecliptic of the Solar system on equinox J2000. Planetary masses, semi-major axes of orbits, eccentricities and arguments of pericentres of choosen exoplanetary systems are given from~\cite{Schneider2017}. Inclinations are assumed to be $5^\circ$ for all orbits. Values of longitudes of the ascending nodes and mean anomalies vary for the estimation of the Hamiltonian. Then the Kepler elements are converted to the Poincare elements and used for the evaluation of the Hamiltonian expansion. 

In the Table~\ref{tab3} the evaluation of the Hamiltonian terms and their approximation accuracy are presented for three chosen planetary systems. Columns which are denoted as $\Delta s$ consist absolute values of relative differences between the evaluated series expansion of the Hamiltonian terms (in previous column) and the evaluated exact expression.

\begin{table}[h]
	\caption{The estimation accuracy of the Hamiltonian for the Solar system's giant-planets and four-planetary systems HD141399, HD160691} \label{tab3}
	\begin{tabular}{c|cc|cc|cc}
	\hline
	Indices & Giant-planets & $\Delta s$ & HD141399 & $\Delta s$ & HD160691 & $\Delta s$ \\
	\hline\hline
	$i,j$ & \multicolumn{6}{c}{The main part of the disturbing function} \\
	\hline
	$1,2$ & $-2.91626\cdot10^{-2}$ & $2\cdot10^{-9}$ & $-1.83560$ & $3\cdot10^{-7}$ & $-1.57429\cdot10^{-2}$ & $1\cdot10^{-7}$ \\
	$2,3$ & $-6.67682\cdot10^{-4}$ & $6\cdot10^{-9}$ & $-0.53022$ & $6\cdot10^{-8}$ & $-3.37462\cdot10^{-1}$ & $3\cdot10^{-6}$ \\
	$3,4$ & $-5.24259\cdot10^{-5}$ & $4\cdot10^{-7}$ & $-0.11329$ & $2\cdot10^{-5}$ & $-6.93740\cdot10^{-1}$ & $4\cdot10^{-7}$ \\
	$1,3$ & $-1.82028\cdot10^{-3}$ & $2\cdot10^{-8}$ & $-0.19286$ & $3\cdot10^{-7}$ & $-3.17282\cdot10^{-2}$ & $4\cdot10^{-6}$ \\
	$2,4$ & $-6.02633\cdot10^{-4}$ & $1\cdot10^{-9}$ & $-0.22267$ & $4\cdot10^{-4}$ & $-1.39172\cdot10^{-1}$ & $6\cdot10^{-7}$ \\
	$1,4$ & $-1.90924\cdot10^{-3}$ & $2\cdot10^{-9}$ & $-0.07089$ & $3\cdot10^{-4}$ & $-9.83227\cdot10^{-3}$ & $6\cdot10^{-7}$ \\
	\hline\hline
	$k$ & \multicolumn{6}{c}{The second part of the disturbing function} \\
	\hline
	$2,3,4$ & $3.22681\cdot10^{-3}$ & $3\cdot10^{-8}$ & $-2.71023$ & $1\cdot10^{-4}$ & $-1.23557\cdot10^{-1}$ & $2\cdot10^{-5}$ \\
	\hline\hline
	& \multicolumn{6}{c}{The Hamiltonian} \\
	\hline
	& $3.22505\cdot10^{-2}$ & $7\cdot10^{-13}$ & $-5.23315\cdot10^{-5}$ & $8\cdot10^{-8}$ & $-3.33545\cdot10^-4$ & $4\cdot10^{-9}$ \\
	\hline
	\end{tabular}
\end{table}

The estimation accuracy for the Solar system's giant planets is the best because their eccentricities are less than for exoplanetary systems and not exceed the value of $0.06$.

At the last step we have constructed semi-analytical motion theory up to the second degree of planetary masses. The averaged Hamiltonian of the problem and the generating function of the transformation between osculating and averaged elements are constructed up to terms with the second degree of the small parameter. Motion equations and functions for the change of variables are given in the second approximation of the Hori--Deprit method.

The number of terms and the computation time of the averaged Hamiltonian and the generating function are shown in Table~\ref{tab4}. The number of terms $N_1$, $N_2$ and the computation time $t$ of the motion equations and the functions for the change of variables for the first and the second orders of the motion theory are given in Table~\ref{tab5} and Table~\ref{tab6}.

\begin{table}[h]
	\caption{The properties of the averaged Hamiltonian and the generating function} \label{tab4}
	\begin{tabular}{lccccc}
		\hline
		& $H_0$ & $H_1$ & $H_2$ & $T_1$ & $T_2$ \\
		\hline
		The number of terms & 4 & $6\;393$ & $381\;534$ & $2\;774\;983$ & $2\;926\;631\;639$ \\
		The computation time & $0^\text{s}$ & $12^\text{m}30^\text{s}$ & $66^\text{h}40^\text{m}$ & $3^\text{m}50^\text{s}$ & $19^\text{h}40^\text{m}$ \\
		\hline
	\end{tabular}
\end{table}

\begin{table}[h]
	\caption{The properties of the averaged motion equations} \label{tab5}
	\begin{tabular}{c|cccc|c|cccc}
	\hline
	el. & $N_1$ & $t$ & $N_2$ & $t$ & el. & $N_1$ & $t$ & $N_2$ & $t$ \\
	\hline
	$L_1$ & \multicolumn{3}{c}{semi-major axes}    & & $\lambda_1$ & $2\;936$ & $0.2^\text{s}$ & $302\;476$ & $47^\text{s}$ \\
	$L_2$ & \multicolumn{3}{c}{are constant}       & & $\lambda_2$ & $3\;452$ & $0.2^\text{s}$ & $361\;534$ & $57^\text{s}$ \\
	$L_3$ & \multicolumn{3}{c}{in averaged theory} & & $\lambda_3$ & $3\;453$ & $0.2^\text{s}$ & $347\;284$ & $55^\text{s}$ \\
	$L_4$ & \multicolumn{3}{c}{therefore $\dot{L}_k=0\;\forall k$} & & $\lambda_4$ & $2\;939$ & $0.2^\text{s}$ & $255\;418$ & $40^\text{s}$ \\
\hline
	$\xi_{1,1},\eta_{1,1}$ & \;\;\;$906$ & $0.05^\text{s}$ & $33\;400$ & $5.3^\text{s}$ & $\xi_{2,1},\eta_{2,1}$ & \;\;\;$911$ & $0.05^\text{s}$ & \;\;$21\;638$ & $3.4^\text{s}$ \\
	$\xi_{1,2},\eta_{1,2}$ &    $1\;063$ & $0.05^\text{s}$ & $42\;322$ & $6.7^\text{s}$ & $\xi_{2,2},\eta_{2,2}$ &    $1\;071$ & $0.05^\text{s}$ & \;\;$27\;679$ & $4.4^\text{s}$ \\
	$\xi_{1,3},\eta_{1,3}$ &    $1\;060$ & $0.05^\text{s}$ & $42\;576$ & $6.8^\text{s}$ & $\xi_{2,3},\eta_{2,3}$ &    $1\;071$ & $0.05^\text{s}$ & \;\;$27\;971$ & $4.5^\text{s}$ \\
	$\xi_{1,4},\eta_{1,4}$ & \;\;\;$897$ & $0.05^\text{s}$ & $33\;754$ & $5.4^\text{s}$ & $\xi_{2,4},\eta_{2,4}$ & \;\;\;$911$ & $0.05^\text{s}$ & \;\;$22\;316$ & $3.5^\text{s}$ \\
	\hline
	\end{tabular}
\end{table}

\begin{table}[h]
	\caption{The properties of the functions for the change of variables} \label{tab6}
	\begin{tabular}{c|cccc|c|cccc}
	\hline
	el. & \hspace{0.4cm}$N_1$ & \hspace{0.2cm}$t$ & \hspace{0.4cm}$N_2$ & \hspace{0.2cm}$t$ & el. & \hspace{0.4cm}$N_1$ & \hspace{0.2cm}$t$ & \hspace{0.4cm}$N_2$ & \hspace{0.2cm}$t$ \\
	\hline
	$L_1$ & $1.16\cdot10^6$ & $64^\text{s}$ & $1.86\cdot10^9$ & $28.5^\text{h}$ & $\lambda_1$ & $2.36\cdot10^6$ & $131^\text{s}$ & $3.50\cdot10^9$ & $54.0^\text{h}$ \\
	$L_2$ & $1.52\cdot10^6$ & $84^\text{s}$ & $2.29\cdot10^9$ & $35.2^\text{h}$ & $\lambda_2$ & $3.09\cdot10^6$ & $172^\text{s}$ & $4.50\cdot10^9$ & $72.0^\text{h}$ \\
	$L_3$ & $1.52\cdot10^6$ & $84^\text{s}$ & $2.30\cdot10^9$ & $35.3^\text{h}$ & $\lambda_3$ & $3.09\cdot10^6$ & $172^\text{s}$ & $4.50\cdot10^9$ & $72.0^\text{h}$ \\
	$L_4$ & $1.16\cdot10^6$ & $64^\text{s}$ & $1.88\cdot10^9$ & $28.8^\text{h}$ & $\lambda_4$ & $2.36\cdot10^6$ & $131^\text{s}$ & $3.50\cdot10^9$ & $54.0^\text{h}$ \\
	\hline
	$\xi_{1,1},\eta_{1,1}$ & $0.49\cdot10^6$ & $27^\text{s}$ & $309\cdot10^6$ & $4.7^\text{h}$ & $\xi_{2,1},\eta_{2,1}$ & $0.45\cdot10^6$ & $25^\text{s}$ & $197\cdot10^6$ & $3.0^\text{h}$ \\
	$\xi_{1,2},\eta_{1,2}$ & $0.64\cdot10^6$ & $36^\text{s}$ & $401\cdot10^6$ & $6.3^\text{h}$ & $\xi_{2,2},\eta_{2,2}$ & $0.58\cdot10^6$ & $32^\text{s}$ & $258\cdot10^6$ & $4.0^\text{h}$ \\
	$\xi_{1,3},\eta_{1,3}$ & $0.64\cdot10^6$ & $36^\text{s}$ & $398\cdot10^6$ & $6.2^\text{h}$ & $\xi_{2,3},\eta_{2,3}$ & $0.58\cdot10^6$ & $32^\text{s}$ & $256\cdot10^6$ & $3.9^\text{h}$ \\
	$\xi_{1,4},\eta_{1,4}$ & $0.49\cdot10^6$ & $27^\text{s}$ & $305\cdot10^6$ & $4.7^\text{h}$ & $\xi_{2,4},\eta_{2,4}$ & $0.45\cdot10^6$ & $25^\text{s}$ & $194\cdot10^6$ & $3.0^\text{h}$ \\
	\hline
	\end{tabular}
\end{table}

\section{The orbital evolution of the Solar system's giant planets}

Finally we have applied our averaged motion theory to the investigation of the orbital evolution of the Solar system's giant planets. The integration of the motion equations in averaged Poincare elements is performed by Everhart method of 15$^\text{th}$ order. The initial conditions of the integration are based on positions and velocities of giant planets w.r.t mean ecliptic and equinox J2000 on date 01.01.2000. The averaged Poincare elements are obtained by using of the functions for the the change of variables in the second approximation. The time interval of the integration is 10~billion years with the time step $10\:000$ years. The motion of the planets has an almost periodic character. Eccentricities and inclinations of the planetary orbits save small values. The short-term perturbations remain small over the entire period of the integration.

\begin{figure}[h!]
	\includegraphics[scale = 0.95]{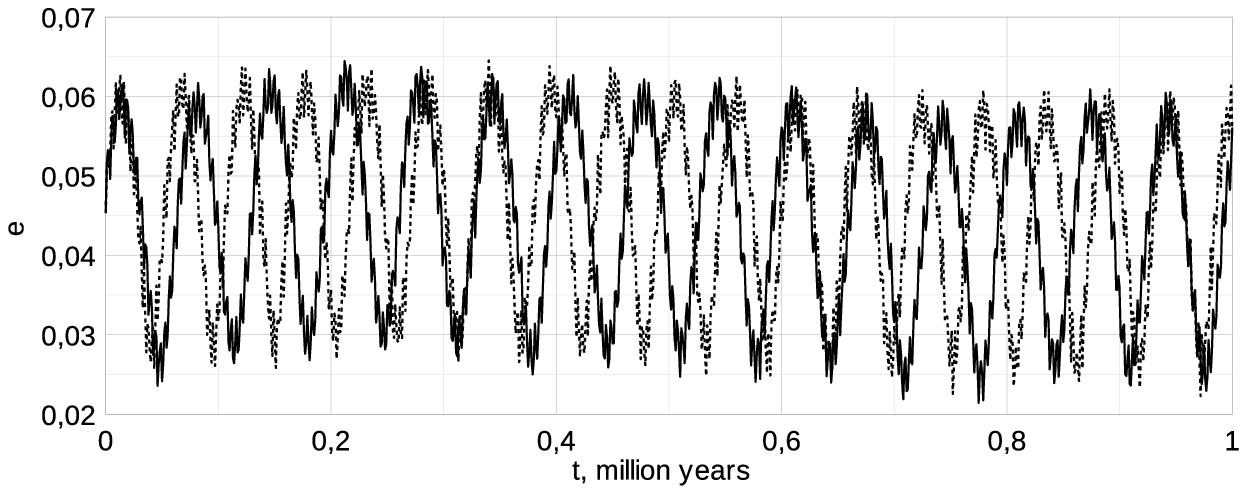}
	\caption{The evolution of the osculating eccentricity of Jupiter orbit over 1~million years. Solid line is data of semi-analytical motion theory, dashed line is data of Cowell-Stormer numerical integration}
	\label{figure1}
\end{figure}

\begin{figure}[h!]
	\includegraphics[scale = 0.95]{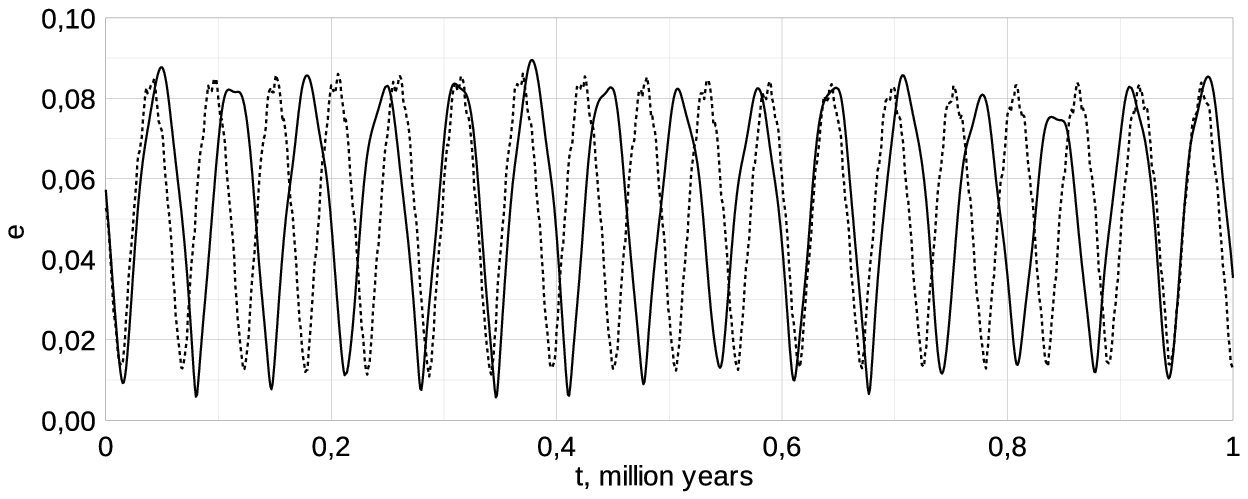}
	\caption{The evolution of the osculating eccentricity of Saturn orbit over 1~million years. Solid line is data of semi-analytical motion theory, dashed line is data of Cowell-Stormer numerical integration}
	\label{figure2}
\end{figure}

\begin{figure}[h!]
	\includegraphics[scale = 0.95]{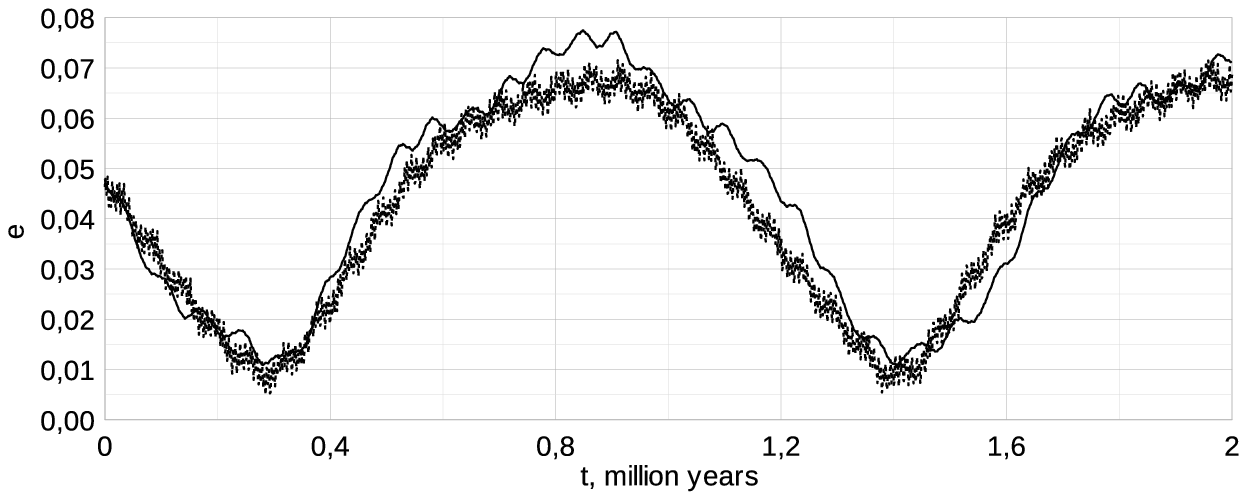}
	\caption{The evolution of the osculating eccentricity of Uranus orbit over 2~million years. Solid line is data of semi-analytical motion theory, dashed line is data of Cowell-Stormer numerical integration}
	\label{figure3}
\end{figure}

\begin{figure}[h!]
	\includegraphics[scale = 0.95]{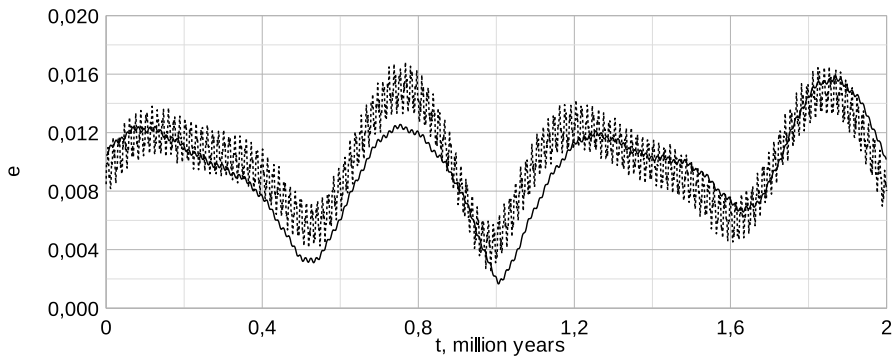}
	\caption{The evolution of the osculating eccentricity of Neptune orbit over 2~million years. Solid line is data of semi-analytical motion theory, dashed line is data of Cowell-Stormer numerical integration}
	\label{figure4}
\end{figure}

We used a modified Cowell-Stormer numerical integrator with modifications by W.~I.~Newman (according to \cite{Goldstein1996}, \cite{Varadi1999}) for the comparison with obtained results of our semi-analytical motion theory. The evolution of osculating eccentricities of Jupiter (in Figure~\ref{figure1}),  Saturn (in Figure~\ref{figure2}), Uranus (in Figure~\ref{figure3}) and Neptune (in Figure~\ref{figure4}) orbits is given in barycentric coordinates. In each figures solid line shows data of semi-analytical motion theory, dashed line shows data of Cowell-Stormer numerical integration. Some properties of orbital eccentricities and inclinations such as their minimum $e_{min}$, $I_{min}$ and maximum $e_{max}$, $I_{max}$ values as well as amplitudes $a_{e}$, $a_{I}$ and periods $T_{e}$, $T_{I}$ of the change are presented in Table~\ref{tab7}. The difference between periods of Jupiter and Saturn orbital eccentricities in numerical and semi-analytical results can be explained by not taking into account some terms of the second approximation. In the rest these results show qualitative agreement with the direct numerical integration.

The terms of the second approximation of the averaged Hamiltonian are constructed up to the first degree of eccentric and oblique Poincare elements. So, in the present work we do not consider the terms of the averaged Hamiltonian that lead to ``the great inequality'' between Jupiter and Saturn. These terms have minimum third degree of eccentric and oblique Poincare elements and ones will take into account in the future work together with the terms of the third approximation of the averaged Hamiltonian. 

\begin{table}[h!]
\caption{The limits of the change of orbital osculating eccentricities and inclinations, their amplitudes and periods in barycentric frame}
\label{tab7}
  \begin{tabular}{l|cc|cc}
    \hline
    \multicolumn{1}{c|}{} & \multicolumn{2}{c}{Jupiter} & \multicolumn{2}{|c}{Saturn} \\\hline
    & \multicolumn{1}{c}{semi-analytical} & \multicolumn{1}{c}{Cowell-Stormer} & \multicolumn{1}{|c}{semi-analytical} & \multicolumn{1}{c}{Cowell-Stormer} \\\hline
    $e_{min}$      & $0.0201$   & $0.0220$   & $0.0041$   & $0.0093$   \\
    $e_{max}$      & $0.0657$   & $0.0647$   & $0.0903$   & $0.0870$   \\
    $a_{e}$        & $0.0228$   & $0.0213$   & $0.0369$   & $0.0389$   \\
    $T_{e}$, years & $66\;225$  & $54\;896$  & $66\;225$  & $54\;896$  \\\hline
    $I_{min}$, $^\circ$ & $1.0953$   & $1.0976$   & $0.5734$   & $0.5656$  \\
    $I_{max}$, $^\circ$ & $2.0591$   & $2.0587$   & $2.5868$   & $2.5850$  \\
    $a_{I}$,   $^\circ$ & $0.4819$   & $0.4805$   & $1.0067$   & $1.0097$  \\
    $T_{I}$, years      & $49\;019$  & $49\;217$  & $49\;019$  & $49\;217$ \\
    \hline\hline
    \multicolumn{1}{c|}{} & \multicolumn{2}{c}{Uranus} & \multicolumn{2}{|c}{Neptune} \\\hline
    & \multicolumn{1}{c}{semi-analytical} & \multicolumn{1}{c}{Cowell-Stormer} & \multicolumn{1}{|c}{semi-analytical} & \multicolumn{1}{c}{Cowell-Stormer} \\\hline
    $e_{min}$      & $0.0053$      & $0.0026$      & $0.0006$               & $0.0020$ \\
    $e_{max}$      & $0.0791$      & $0.0736$      & $0.0161$               & $0.0168$ \\
    $a_{e}$        & $0.0369$      & $0.0355$      & $0.0077$               & $0.0074$ \\
    $T_{e}$, years & $1\;111\;000$ & $1\;110\;000$ & $370\;000$, $556\;000$ & $357\;000$, $526\;000$ \\\hline
    $I_{min}$, $^\circ$ & $0.4404$   & $0.4065$   & $0.7883$      & $0.7935$ \\
    $I_{max}$, $^\circ$ & $2.7116$   & $2.7689$   & $2.3728$      & $2.3782$ \\
    $a_{I}$,   $^\circ$ & $1.1356$   & $1.1812$   & $0.7922$      & $0.7924$ \\
    $T_{I}$, years      & $435\;000$ & $434\;000$ & $2\;000\;000$ & $1\;998\;000$ \\\hline
  \end{tabular}
\end{table}

\section{Conclusion}

The expansion of the Hamiltonian of four-planetary system into the Poisson series is constructed up to 2$^\text{nd}$ degree of the small parameter. The estimation accuracy of the Hamiltonian expansion is about $10^{-12}$ for the Solar system's giant planets and it is about $10^{-8}$ for chosen extrasolar planetary systems.

We have constructed the averaged Hamiltonian and the generating function of the transformation up to 2$^\text{nd}$ degree of the small parameter. The motion equations and the functions for the change of variables were constructed in 2$^\text{nd}$ approximation too.

We have applied the constructed semi-analytical motion theory to the investigation of orbital evolution of the Solar system's giant planets. The motion of the planets has an almost periodic character. The orbital eccentricities and inclinations save small values over the entire period of the integration, which is equal to 10~billion years. The results of the integration show qualitative agreement with the direct numerical integration.

In the process of our calculations Piranha shows the ability to work with very large series.

In the future our semi-analytical motion theory will be used for the investigation of dynamic properties of various extrasolar planetary systems.


\subsection*{Acknowledgment}
We are grateful to anonymous reviewers whose constructive and valuable comments greatly helped us to improve the paper.

This work is funded by RFBR according to the research project no. 18-32-00283 and the Government of the Russian Federation (Act no. 211, agreement no. 02.A03.21.0006).

\end{document}